\documentclass{aa}
\usepackage{natbib}
\bibpunct{(}{)}{;}{a}{}{,} 

\usepackage{graphicx}
\usepackage{txfonts}
\usepackage{xcolor}
\usepackage{lmodern}
\usepackage{caption}
\usepackage{subcaption}
\usepackage{mathtools}
\usepackage{indentfirst}
\usepackage{makecell}
\usepackage{orcidlink}
\usepackage{hyperref}
\hypersetup{
    colorlinks=true,
    linkcolor=blue,
    citecolor=blue}

\newcommand{\boldchanges}[1]{#1}

\newcommand{\SIXTE}[0]{\texttt{SIXTE}}
\defcitealias{audard_bulk_2025}{XRISM Collaboration et al. 2025a}
\defcitealias{xrism_collaboration_2025_abell_2029_a}{XRISM Collaboration et al. 2025b}
\defcitealias{xrism_collaboration_xrism_2025}{XRISM Collaboration et al. 2025c}

\begin{document} 

\titlerunning{Toward mapping turbulence in the intracluster medium IV. }
   \title{Toward mapping turbulence in the intracluster medium}
   \subtitle{IV. Using NewAthena/X-IFU and simulation-based inference to constrain turbulence}


   \author{A. Molin\orcidlink{0009-0007-9783-3752} 
          \inst{1},
          S. Dupourqué\orcidlink{0000-0003-2715-8986}
          \inst{1},
          N. Clerc
          \inst{1},
          E. Pointecouteau
          \inst{1},
          F. Pajot
          \inst{1},
    E. Cucchetti\orcidlink{0000-0002-5548-4351}
          \inst{2}
          }

\authorrunning{A. Molin}

   \institute{IRAP, Université de Toulouse, CNRS, CNES, UT3-PS, Av. du Colonel Roche 9, 31400, Toulouse, France
             \and
    Centre National d’Etudes Spatiales, Centre spatial de Toulouse, 18 avenue Edouard Belin, 31401 Toulouse Cedex 9, France}


 
  \abstract
   {The NewAthena mission planned for launch in the late 2030s will carry X-IFU, an integral field unit spectrometer that will obtain unique insights into the X-ray hot Universe through its combination of spectral and spatial capabilities. \boldchanges{Its ability to deliver spatially resolved observations with a high spectral resolution will allow for detailed mapping of turbulent velocities in the hot gas of galaxy clusters, offering capabilities beyond those of previous missions to study their complex dynamics.}} 
   {This is the fourth in a series of papers aimed at forecasting the ability to investigate  turbulence in the intracluster medium through the observation of the centroid shift caused by turbulent motions of the gas.  In this paper we improve on previous methods by investigating the ability of simulation-based inference (SBI) to constrain the underlying nature of velocity fluctuations through the use of standard observational diagnostics, such as the structure function.}
   {We rely on a complex architecture of neural networks including masked autoencoders in order to model the likelihood and posterior distributions relevant to our case. We investigate its capability to retrieve the turbulence parameters on mock observations, and explore its capability to use alternative summary statistics.}
   {Our trained models are able to infer the parameters of the intracluster gas velocity power spectrum in independently simulated X-IFU observations of a galaxy cluster. We evaluated the precision of the recovery for different models. We show the necessity to use methods such as SBI to avoid an underestimation of the sources of variance by comparing the results to our previous paper. We confirm that cluster-to-cluster variations in the stochastic velocity field (i.e., sample variance) severely impact the precision of recovered turbulent features. Our results demonstrate the need for advanced modeling methods to tackle the complexity of the physical information nested within future high-resolution spectroscopic X-ray observations, such as the ones expected from X-IFU/NewAthena.}
   {}


   \maketitle

\section{Introduction}
\indent Galaxy clusters are the end result of the hierarchical structure assembly of matter, making them the largest gravitationally bound objects. They are permeated by the intracluster medium (ICM), which is a highly ionized plasma that accounts for 80\% of their baryonic mass content. 
Turbulence within the ICM is believed to play an important role in dissipating the energy injected by a variety of gravitational and nongravitational processes. These processes include feedback from active galactic nuclei (AGNs) \citep{mcnamara_mechanical_2012, eckert_feedback_2021, voit_global_2017}, merger events and accretion \citep{zuhone_merger_2022,voit_role_2018, nelson_evolution_2012}, either with sub-halos and galaxies, other clusters, or by the continuous accretion of matter from the surrounding medium and filaments of the cosmic web. Turbulence is expected to contribute to the nonthermal pressure support of galaxy clusters and thus it needs to be accounted for when inferring the total mass of galaxy clusters from the distribution of intracluster gas \citep{2016ApJ...827..112B, voit_role_2018, 2019SSRv..215...25P}. \\
\indent Studying turbulence within galaxy clusters allows one to get an insight into the mechanisms driving the dynamics of the ICM, and is a way to estimate the fraction of nonthermal support \citep{vazza_turbulent_2018, angelinelli_turbulent_2020}. \\
\indent Previous studies in the X-ray have used surface brightness fluctuations to constrain turbulence in clusters \citep[e.g.,][]{schuecker_probing_2004, Churazov2012, Zhuravleva2012,Zhuravleva2018,Simionescu2019,dupourque_investigating_2023,dupourque_chex-mate_2024}. This technique requires an assumption about the relation between density fluctuations and the underlying turbulent motions of the gas. \\
\indent The ICM is an optically thin medium in the X-ray. Any movement within the gas is traced as a spectral distortion (e.g., broadening, Doppler shifts) of the emitted spectrum, and in particular the emission lines of the heavy elements \citep{inogamov_turbulence_2003}. Direct observations of turbulence become possible through the study of line centroid shifts and broadening. Probing turbulent motion requires a high spectral resolution. Current imaging spectrometers (such as EPIC on board \textit{XMM-Newton} or ACIS on board \textit{Chandra}) enable at best $100-150$~eV resolution, corresponding to a resolving power of R\,$\sim$\,10. Such resolving power is insufficient to capture spectral line distortions acting at a level of, for example, $\sim100$\,km/s. Instruments relying on dispersive spectroscopy, such as grating-based spectrometers (e.g., RGS on board \textit{XMM-Newton}) have a greater resolution but suffer from mixing of spatial scales \citep[see e.g.,][for upper constraints on turbulence using RGS, with a sample of clusters]{pinto_chemical_2015, sanders_constraints_2011}. Only recently have arrays of microcalorimeters had a sufficient spectral and spatial resolution to deliver the first maps of ICM motions, with observations using SXS on board Hitomi \citep{hitomi_collaboration_quiescent_2016,aharonian_hitomi_2018} and its copy Resolve on board XRISM \citepalias{audard_bulk_2025, xrism_collaboration_xrism_2025, xrism_collaboration_2025_abell_2029_a}.

\indent The NewAthena space telescope is a planned mission of the European Space Agency, to be launched in the mid-2030s. The mission is destined for the observation of the hot and energetic Universe in X-ray between 0.2 and 12~keV \citep{2013arXiv1306.2307N}. Relying on silicon pore optics for its mirror \citep{10.1117/12.3021135}, NewAthena will combine a high collecting area, with 1.2 $\text{m}^2$ at 1 keV, and a spatial resolution twice that of XMM-Newton, with a point spread function (PSF) of 8-9 arcseconds half-equivalent width. NewAthena will embark with two instruments. The Wide Field Imager, using DEPFET sensors, will allow wide-field imaging at a high pixel density \cite{10.1117/12.3019707}. The X-IFU will consist of a hexagonal array of 1504 transition edge sensors, each providing a resolution of 4 eV or less \citep[with 3eV currently considered the goal][]{2023ExA....55..373B}.  The spectral and spatial resolutions of X-IFU will allow us to get an unprecedented insight into the motions of the ICM with the spatial mapping of X-ray emission line shifts and broadening. The NewAthena space telescope, and hence the X-IFU instrument, has undergone a phase of important reformulation since our previous paper, leading to a substantially different mirror and readout design. The results presented in this paper incorporate these latest updates to the X-IFU performance specifications (unless otherwise stated). \\
\indent Addressing the wide and complex topic of turbulence is an observationally challenging task. A first-order approach involves constraining simplified parametric models capturing the main features of the random motions in the hot baryonic gas. In return, these models can then predict the nonthermal pressure support and the correction to hydrostatic equilibrium. It is commonly assumed that all turbulence within the cluster can be described by an isotropic velocity power spectrum. Assuming that the turbulent field can be described by a Gaussian random field, this power spectrum is parametrized by a handful of physical parameters (slope, amplitude, etc.). The parameters of this power spectrum are then the parameters to recover through observations.\\
\indent The 2D structure function derived from a reconstructed (projected) velocity map is a statistical description of the velocity map. It encapsulates the amplitude of fluctuations at different scales and it is correlated with the physical parameters underlying the power spectrum of the velocity field \citep[e.g.,][]{Zhang_LEM_2024, xrism_collaboration_xrism_2025}. The analytical prediction of the exact shape and in particular the exact error budget of the structure function for a given observation setup (i.e., binning map, emissivity model) is a highly challenging task because of the nature of this problem. In context, there is no analytical formulation for the likelihood of the 2D structure function on a binned velocity map. Traditional inference methods, such as gradient descent or Monte Carlo Markov chain sampling, require an assumption about the form of the likelihood. Accurately estimating the associated error budget requires one to account for all possible sources of variance. In particular, sample variance, or cosmic variance as coined by \cite{ZuHone2016c}, describes the fact that each turbulent random field is only a single finite-size realization among an infinity with the same underlying power spectrum.\\
\indent In our first two papers \citep{clerc_towards_2019, cucchetti_towards_2019}, we focused first on analytical cases, analyzing the 2D structure function of velocity maps and its associated sources of variance and errors. With the third paper \citep[][B24 hereafter]{2024A&A...686A..41B}, we applied these developed capabilities of predicting accurately the structure function and its errors to the case of mock X-IFU observations. Working from an idealized observational setup, we assessed the constraints that could be put on the turbulent spectrum. We assumed the form of the likelihood to be Gaussian,  with Gaussian errors on the structure function and zero covariance. Results were shown for mock observations of 19 adjacent pointings on a $z=0.1$ cluster, mapping the cluster out to $R_{500}$, assuming an unrealistically large total exposure time and ignoring the effect of astrophysical and instrumental backgrounds.\\
\indent In this paper, we leverage simulation-based inference (SBI hereafter) through the \texttt{sbi} Python package \citep{tejero-cantero2020sbi} to infer the parameters of the power spectrum with X-IFU observations of a galaxy cluster. Simulation-based inference relies on a neural network, optimized for this task. The network learns the relation between the observables and the parameters used to generate them, fed with a large number of simulations giving pairs of parameters, $\theta_i$, and observations, $X_i$. Since it is trained with simulations that include the stochasticity of random Gaussian fields, this method naturally accounts for this source of variance. The need for \boldchanges{cumbersome and} computationally heavy analytical computations is then less of a concern. We use this new method to investigate mock observations identical to our previous paper. This method has successfully been used in astrophysics (see e.g., pulsar population synthesis \cite{2024ApJ...968...16G}, weak lensing analysis \cite{2023MNRAS.524.6167L}, X-ray spectra fitting \citep{2024A&A...686A.133B}) as well as for constraining turbulence from surface brightness fluctuation maps in \cite{dupourque_investigating_2023, dupourque_chex-mate_2024}. In addition, if one is able to provide different observable properties arising from the same process, it is straightforward to add them to the inference with SBI. Moreover, the problem of dealing with the sample variance can be seen as an issue of marginalizing the likelihood over all possible realizations of the turbulent velocity field. Simulation-based inference is naturally suited for tackling marginalization over nuisance parameters, which further motivates its use in this problem. \\
\indent The paper is organized in the following way. In Section~\ref{sec:2} we introduce the method used for X-IFU/NewAthena mock observations. In Section~\ref{sec:3} we introduce SBI, and the model used for training, and we compare the results with our previous paper. In Section~\ref{sec:results} we use this new approach on the new X-IFU configuration. We look into the impact of adding astrophysical and instrumental  background. We also investigate the effectiveness of adding further summary statistics of the observables for the constraining power of our SBI base method. In Section~\ref{sec:discusssion} we provide conclusions on this new data analysis strategy, and on our inferred capabilities to derive constraints on the turbulence of the ICM with the X-IFU instrument.\\
\indent This study has been conducted with the assumption of a $\Lambda$CDM cosmology, with $H_0=70$\,km\,s$^{-1}$\,Mpc$^{-1}$, $\Omega_m=0.3$, and $\Omega_\Lambda=0.7$. With such parameters, at a redshift of z = 0.1, 1 kiloparsec (kpc) corresponds to an angular extent of 0.54 arcseconds (arcsec)\boldchanges{, and one pixel has a projected size of 10~kpc}. \\
\indent All corner plots shown in this paper show the 1 and 2$\sigma$ contours as shaded regions. The marginalized distributions in the diagonal show the 1$\sigma$ region as shaded.

\section{Simulated X-IFU/NewAthena observations}
\label{sec:2}
\indent The following section introduces the tools and models used to produce realistic mock X-IFU observations. As this work relies heavily on the simulations introduced in \boldchanges{B24}, we invite the reader to refer to this paper for a full description, and provide only the essentials below.

\subsection{Cluster toy model}

\indent The X-IFU field of view (FoV) and the NewAthena telescope PSF constrain the maximal and minimal spatial scales observable in a single pointing. Their values are, respectively, $\sim$~4 arcmin and $\sim$~9 arcsec. In this study we chose a cluster at a redshift of $z = 0.1$ with a physical characteristic scale  of $R_{500} = 1300$ kpc, following our previous paper. This size allows us to grasp the cluster entirely within a few pointings in radius. $R_{500}$ is the radius at which the mean density of the cluster is 500 times the critical density of the Universe at this redshift. At that redshift, the size of the PSF and the FoV correspond approximately to 18 kpc and 490 kpc, respectively. This allows one to grasp most of the inertial range of turbulence within this medium, which is believed to range from a few hundreds of kiloparsecs down to a few kiloparsecs or even sub-kiloparsec scales. \\
\indent We assume spherical symmetry for simplicity. The density model and temperature profile models are taken from \cite{Ghirardini2019}. The metallicity profile is taken from \cite{Mernier2017}, and the abundances are taken from \citet{AndersAndGrevesse89}. The exact parametrization of profiles can be found in Appendix \ref{AppendixA}.\\
\indent Finally, the emission model is the astrophysical plasma emission code \citep[\texttt{bapec}][]{APECPaper} model in \texttt{xspec} \citep[][\texttt{apec v12.13.0c}]{XSPECPaper}. This model includes the continuum, dominated by Bremsstrahlung emission, as well as all the emission lines for elements up to zinc in a optically thin and collisional plasma at equilibrium, with a thermal broadening. An absorption component (\texttt{phabs} model under \texttt{xspec}) multiplies the emission and is representative of the absorption by hydrogen along the line of sight. The hydrogen column density is fixed to a value of $n_H = 3 \cdot 10^{20}\text{ cm}^{-2}$. \\
\indent \boldchanges{Cluster properties were computed on a discrete 3D grid. For computational issues, we fixed the size of a cell in physical units to less than half of the PSF (i.e., 5~kpc). The grid extended to $+/- 10.8\times R_{500}$ along the line of sight and $+/- 1.4\times R_{500}$ on the plane of the sky, respectively. The cluster properties were then coupled at the level of each grid cell to the emission model through a random sampling of photons in the corresponding cell spectrum. Stacked over the whole grid, we obtained a photon list emitted by the cluster for the given exposure time. This photon list served as an input to the \texttt{SIXTE} simulator described in Sec.~\ref{sec:instr_model}. We strictly followed the procedure by \citet{cucchetti_athena_2018}, to which we refer the reader for a detailed description.}  

\subsection{Turbulent velocity}

\indent In order to model turbulence in the cluster, we assumed a single isotropic and homogeneous power spectrum. Our chosen parametrization follows that of \cite{ZuHone2016c, dupourque_investigating_2023}, and is written as
\begin{equation}
\label{P3Deqn}
    P_{\mathrm{3D}}(k) = \sigma^2 \frac{\displaystyle k^{-\alpha} e^{-(k_{\text{inj}}/k)^2} e^{-(k/k_{\text{diss}})^2}}{\displaystyle \int 4 \pi  k^2 k^{-\alpha} e^{-(k_{\text{inj}}/k)^2} e^{-(k/k_{\text{diss}})^2} dk}
,\end{equation}
where $k$ is the norm of the 3D wavenumber, $k_\text{inj}, k_\text{diss}$  are the inverse of the injection and dissipation scales ($L_\text{inj}, L_\text{diss}$), respectively, $\alpha$ is the slope of the cascade, and $\sigma$ the norm of the spectrum, corresponding to the total dispersion of the 3D speed. The norm is set such that the 3D integral of the power spectrum gives $\sigma^2$.  Similarly to B24, we kept identical values for these parameters, 300 kpc for the injection scale, 10 kpc for the dissipation scale, -11/3 for the slope, and 250 km/s for the norm. This cube of turbulent speed is used as the input for the redshift of the simulated \boldchanges{grid of emission properties} within the cluster. \boldchanges{We thus computed a 3D grid of turbulent speed used together with other physical properties to compute the cluster emission.}\\

\subsection{Background and foreground emissions} 
\label{sec:bkg}

\indent The astrophysical foreground and background emissions in our simulations are accounted for by following the model proposed by \citet{2014A&A...569A..54L}. There are three components, which are the unabsorbed thermal model representing the local bubble, the absorbed one for the Galactic halo, and the absorbed power law for the cosmic X-ray background (CXB). They are modeled under XSPEC as \texttt{apec}, \texttt{phabs*apec}, and \texttt{phabs*powerlaw}, respectively. The parametrization of these models is provided by \citet{2014A&A...569A..54L}. The hydrogen column density is kept at the same value as for the cluster model.  We assume that these diffuse components have no spatial variation across the X-IFU FoV. 

\indent We also accounted for the instrumental background in our simulations. The \SIXTE~ tool, described in the next section, allows one to model the instrumental background, which is set according to the X-IFU requirements of $5\times10^{-3} \mathrm{counts/s/cm}^2\mathrm{/keV}$ \citep{peille_x-ray_2025}. The main source of instrumental background is the high-energy cosmic rays hitting the instrument in the neighborhood of the detectors, and it dominates the background budget at high energies (beyond $\sim 4$ keV). The instrumental background has been updated to a value of $8\times10^{-3} \mathrm{counts/s/cm}^2\mathrm{/keV}$ during the writing of this work; however, we assessed that this change only has a negligible impact on the measurement error described in Sec~\ref{sec:fast_model}. 

\subsection{Instrument model and post-processing}
\label{sec:instr_model}

\indent For a complete model of the X-IFU instrument, we relied on the \SIXTE~ simulator \citep{wilms2014athena, SIXTEPaper}. This simulator takes as inputs instrument configuration files and a photon list, and produces a list of events for the given observation time and instrument attitude. It accounts for the entire instrument, from the mirrors to the electronics, resulting in a high-fidelity simulation of X-IFU observations. Throughout all of this paper, we use a mosaic of 19 X-IFU pointings, with 125~ks of observation for each pointing. \boldchanges{This configuration is the same as the one presented in B24. It was chosen as a simple demonstration case of an ideal low statistical error observation. The latter is ensured by the 125~ks per pointing, a value close to the ad hoc 100~ks commonly adopted in the initial Athena's white paper \citep{nandra_athena_2013} and supporting papers \citep{pointecouteau_hot_2013, ettori_hot_2013}. In B24 and the present work, this configuration allowed for a mapping of the cluster up to $\sim 1$ and $\sim 0.7\, R_{500}$ with a FoV of 5' and 4' equivalent diameter of the pre and post-reformulation of the X-IFU performance parameters by ESA, respectively.} 

\indent The \SIXTE~ simulator outputs a mock event list of the X-IFU observation. The events are binned in regions on the detector array by using Voronoi binning \citep{2003MNRAS.342..345C} to produce bins of an equal signal-to-noise ratio of 200, with this ratio defined as the square root of the total number of counts in each bin. \boldchanges{This produces bins of $\sim10$ pixels at the center and $\sim1000$ pixels on the outskirts}. A spectrum is produced in each bin and fit with the spectral model for the emission of the cluster \boldchanges{(\texttt{bapec} under XSPEC)}, together with the model for the background \boldchanges{(see the above section)}. \boldchanges{The fitting procedure is carried over the 0.2 to 12~keV energy band, without priors on the free parameters: the cluster model norm, temperature, redshift (line centroid), line broadening, metallicity, and background model norm.} 

\begin{figure}
    \centering
    \includegraphics[width=1.0\linewidth]{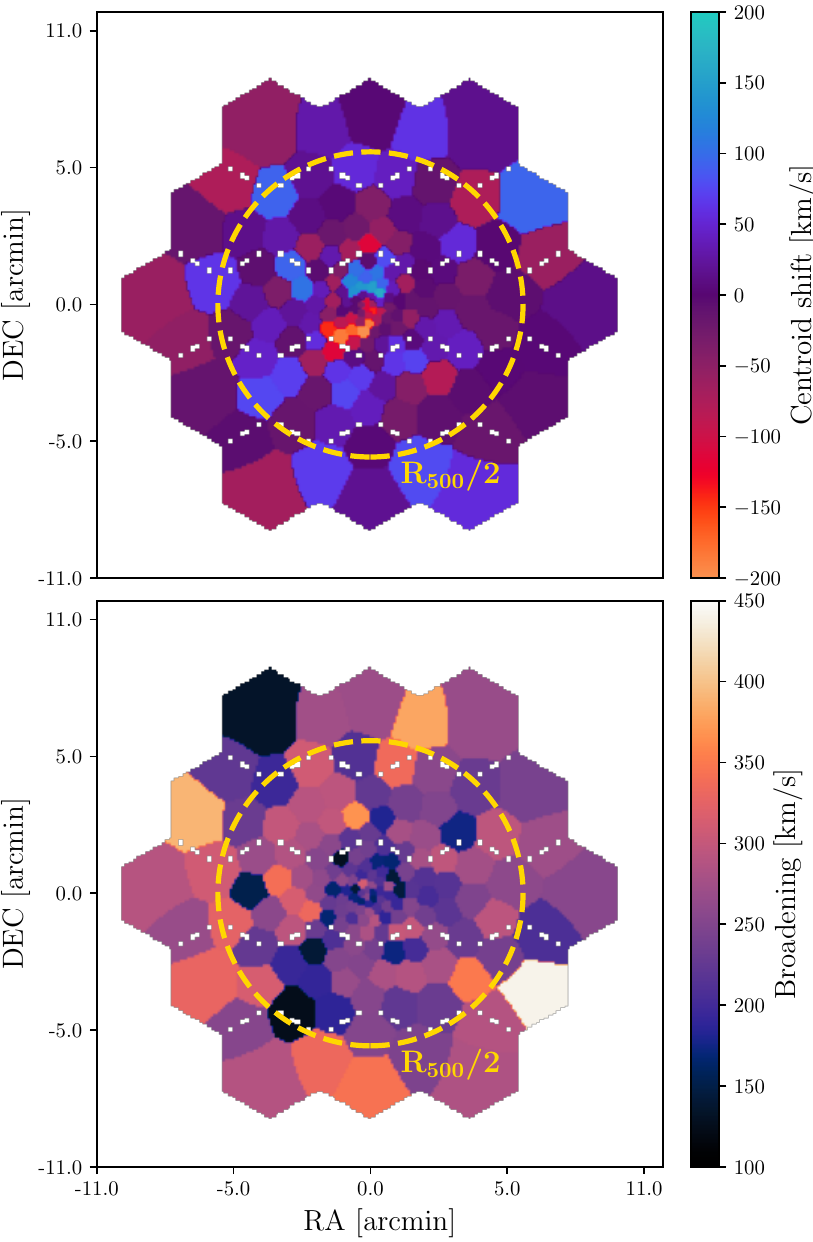}
    \caption{Centroid shift (top) and line broadening (bottom) maps extracted from a mock observation of 19 X-IFU pointings simulated with \SIXTE, centered on a galaxy cluster with a turbulent ICM. The $R_{500}/2$ size of the cluster is shown as a circle on the maps.}
    \label{fig:output_maps}
\end{figure}

For each physical parameter of the fit for the cluster emission, we produce a map where we assign the best fit value in each bin. \boldchanges{The norm, temperature, and abundance of the \texttt{bapec} model are recovered within $\sim1$\% of their input values. } Two parameters of the \texttt{bapec} are of particular interest here: firstly, the velocity, which is the Doppler shifting of the emission lines of the medium \boldchanges{(deduced from the best fit redshift value)}; and secondly, the velocity broadening that results from the Doppler broadening of the emission lines of the medium. Both are quantities integrated along the line of sight. This allows for the production of velocity and broadening maps, which can then be used directly or indirectly as observable quantities to constrain the ICM's turbulence. An example of such maps is shown in Fig.~\ref{fig:output_maps}

\begin{figure}
    \centering
    \includegraphics[width=1.0\linewidth]{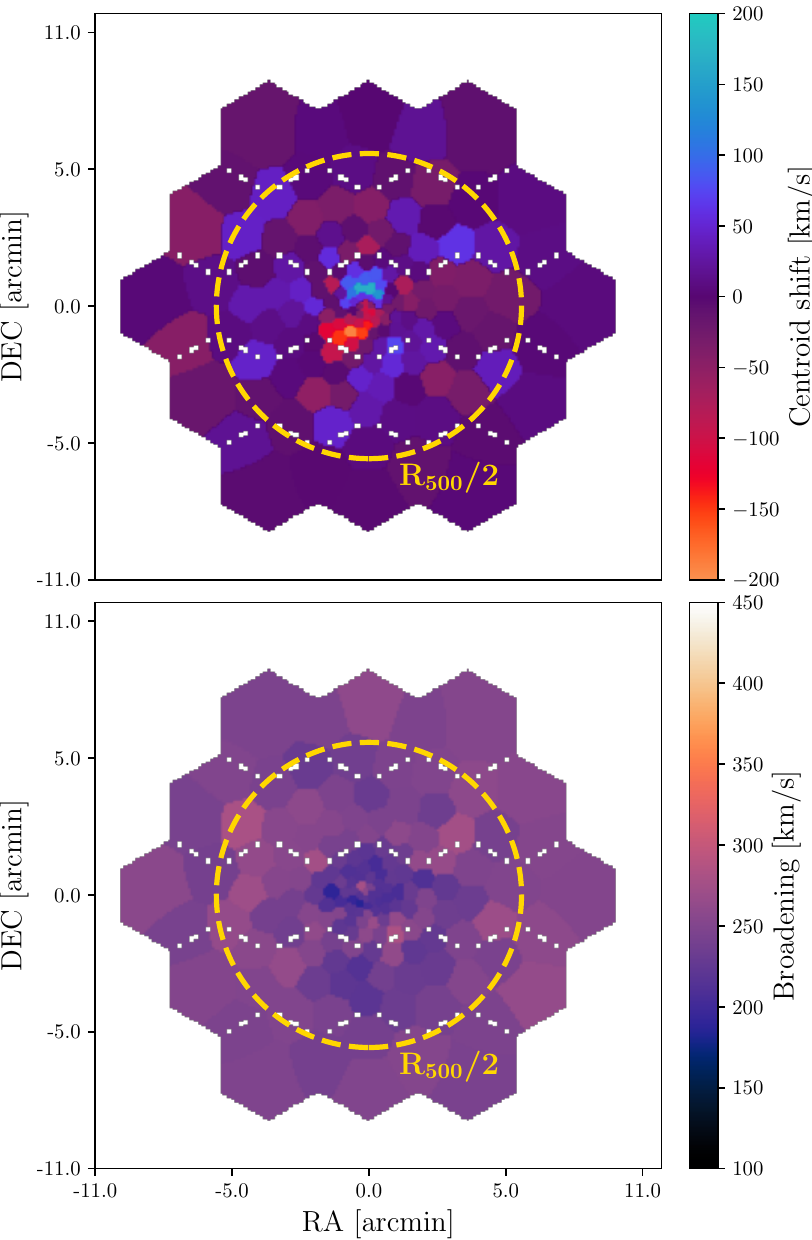}
    \caption{Similar figure as Fig. \ref{fig:output_maps}, showing expected values (inputs maps, see text) of centroid shift and line broadening free of any instrumental noise.}
    \label{fig:input_maps}
\end{figure}

\indent For each observation simulation, we also produce a set of maps, called input maps, which are the emission-weighted projections of the input quantities of the simulation. These input maps are binned similarly to the output maps and are therefore directly comparable. The corresponding input maps to those shown in Fig.~\ref{fig:output_maps} are shown in Fig.~\ref{fig:input_maps}. 

\subsection{Structure function}

\indent One way to summarize the velocity or broadening information is the 2D structure function. Assuming a binned map of the quantity, $C$, the 2D structure function of this map was computed as
\begin{equation}
    \mathrm{SF}(s) = \frac{1}{N_p(s)} \sum_{d(\mathcal{W}, \mathcal{W'}) = s} |C_\mathcal{W} - C_\mathcal{W'}|^2
,\end{equation}
where s is a given separation, $\mathcal{W}$ is a given bin, $d(\mathcal{W}, \mathcal{W'})$ is the distance between the centers of two bins, $C_\mathcal{W}$ is the value of the observed quantity in the given bin, and $N_p(s)$ is the number of pairs of bins that satisfy this separation distance. It is closely related to the 2D power spectrum as it gives out the average power contained in the fluctuations of the mapped quantity at different scales. \\
\indent \boldchanges{The structure function depends on the spatial binning that is used. For the same velocity field projected through emission weighting, different binning schemes will (marginally) affect the shape and normalization of the structure function, and more specifically its average value and variance. As is illustrated in B24, folding the shape and surface of the Voronoi cells in the theoretical computation of the structure function is nontrivial. However, we properly account for the chosen binning scheme by reproducing it in the modeling process of the structure function. Throughout our study, we keep the spatial binning defined for the mock reference observation constant for all simulations generated to train the neural network in our SBI process (see next section). In this way, the binning scheme is properly accounted for in the inference process.}

\subsection{Reference realization}

\begin{figure}
    \centering
    \includegraphics[width=1.0\linewidth]{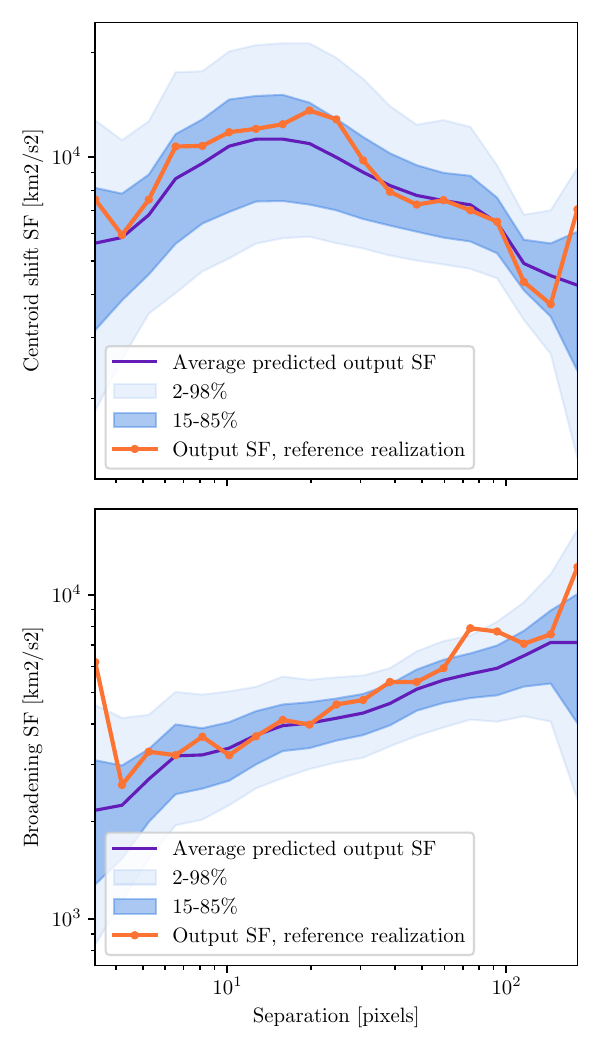}
    \caption{Centroid shift (top) and line broadening (bottom) structure functions extracted from an observation of 19 X-IFU pointings simulated with \SIXTE, with the predicted structure function distributions as percentile envelopes.}
    \label{fig:pred_sf_dist}
\end{figure}

\indent Throughout this paper, we present results on a single realization of a turbulent field within a cluster. This was simulated with \SIXTE, and chosen as a reference. Figure~\ref{fig:pred_sf_dist} shows the structure function of this reference realization. The predicted distribution of structure functions for the true input parameters are also displayed. This realization was chosen as reference because its structure function is close to the average of the predicted distribution. The following results were reproduced with two mock observations of two realizations, with structure functions lying above and below the average. These are presented in Appendix~\ref{AppendixB}. 

\section{Simulation-based inference} 
\label{sec:3}

\subsection{Streamlined model of turbulent field observations}
\label{sec:fast_model}

\indent Simulation-based inference relies on training what is called a neural density estimator, a form of neural network relying on normalizing flows. It is optimized for reproducing the distribution of observables from the distribution of input parameters of a training sample.  In order to train each density estimator, it is necessary to provide a large amount of simulated observations, on the order of $10^5$. To achieve that, we could not rely on the \SIXTE~ simulator as it is too computationally heavy to allow the production of such a number of mock observations. We thus implemented a practical model of turbulent field observations and the reconstruction of structure functions. 

\indent For each mock observation generated by \SIXTE, we have to generate a training set for the neural network to train on, because the observable quantities depend on the features of the observation, such as the binning and the measurement error, which may vary from one observation to the other. The training set is composed of an ensemble of parameters and associated observations $\{\theta_i, X_i\}$. For each realization of parameters, $\theta_i$, we generated a Gaussian random field from the corresponding power spectrum, $\mathcal{P}_{3D}(\theta_i)$, on a fixed grid. We used emission weighting to project the grid on the X-IFU pixel array, and convolved this projection with the instrument PSF. Then, we averaged the velocity obtained in each bin, following the binning scheme of the \SIXTE mock observation, again weighting by the emission, to produce a centroid shift map. Doing the same with a weighted standard deviation of the velocity allowed us to produce a line broadening map. Finally, the measurement error of the centroid shift was added as a Gaussian random variable to the value of each bin. The observable quantity of interest, $X_i$, was then either the structure functions of the output maps or some other summary statistic that we could compute from the maps. In order to produce such a large amount of training samples, we relied on the \texttt{JAX} \citep{frostig_compiling_2019} library, allowing the computation to be run on GPU, which drastically decreased the computation time, with only 5 hours required to produce 500\,000 simulations\footnote{on an NVIDIA Volta V100 GPU}.

\indent The transverse size of the velocity cubes is driven by the size of the X-IFU mosaic. In order to make a comparison with our previous paper, we generated mosaics of 19 pointings, corresponding to a size of 232 X-IFU pixels in the dimension perpendicular to the line of sight. Along the line of sight, the size of the velocity cube is driven by the need to sample the cluster out to $\sim$2.5 $R_{500}$. The size of the cubes is then $232\times232\times1160$ pixels. Because of memory limitations on GPU, the sampling is kept to one point per X-IFU pixel. This sets the smallest scale in the simulated velocity cubes to exactly the adopted dissipation scale (i.e., 10~kpc). 

\subsection{Algorithm and priors}

\indent We require a fast drawing of samples from the posterior distributions in order to alleviate the need to leverage a complete sampler such as MCMC or NUTS for each posterior distribution. The SNPE\_C (that we call SNPE in the remainder of the paper) model \citep{greenberg_automatic_2019} allows just that. In this algorithm, the trained network learns the mapping between the distribution of the input parameters and the distribution of the observable quantities. An independent SNPE density estimator was trained for each choice of summary statistic (see Sec.~\ref{sec:results}). For each trained estimator, we generated 500 000 training samples, with the parameters sampled from uniform distributions such that $\alpha \sim  \mathcal{U}(0.1,8.8)$, $\log_{10}(L_{inj} [\mathrm{kpc}] )\sim \mathcal{U}(2,3)$ and $\sigma \sim \mathcal{U}(50,500) [\mathrm{km/s}]$. We do not expect the constraints on the slope to be tight and left a large prior. The upper edge at 8.8 prevents slopes that are too steep and that cause numerical issues in the power spectrum and Gaussian random field generation. 

\subsection{Modeling the measurement error}

\indent To estimate the measurement error that is made for the reference realization, we simulated ten different draws of the photons input to \SIXTE. \boldchanges{Each mock observation requires the generation and processing of 19 X-IFU pointings through the \texttt{SIXTE} simulator. For computational aspects, we had to limit ourselves to ten draws, though this limited number only provides a crude estimate of the measurement error. However, a more general method of estimating the measurement error that does not involve generating multiple draws for each single realization of a velocity field would be preferred for generalization purposes.}

\begin{figure}
    \centering
    \includegraphics[width=1.0\linewidth]{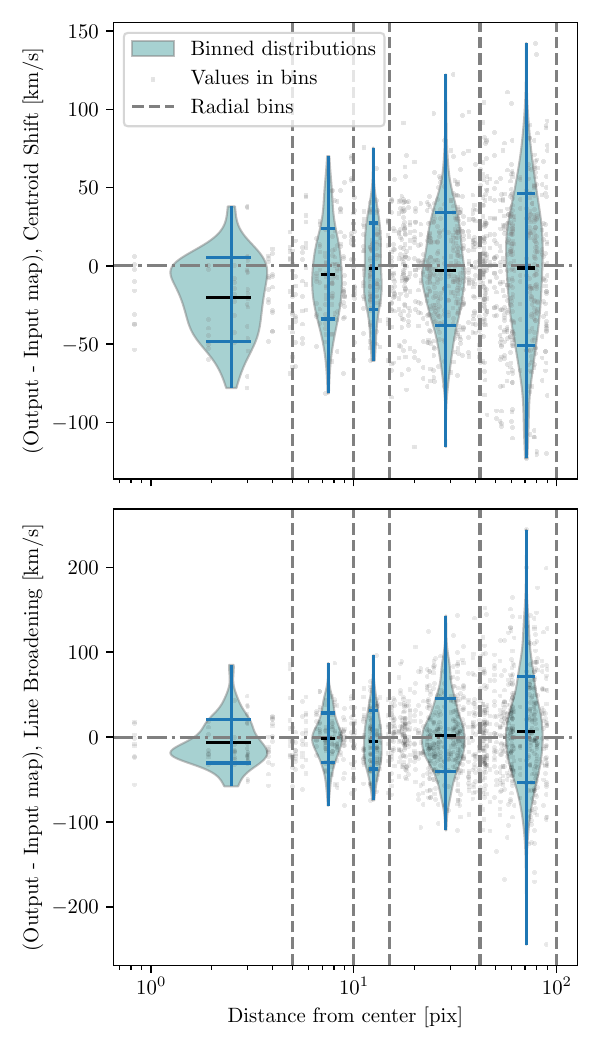}
    \caption{Scatter plots of the binned differences between the output and input maps of centroid shift and broadening, as a function of the distance from the center of the cluster. The different radial bins used for modeling this difference are plotted as vertical lines. The distribution in each radial bin is shown as a violin plot, \boldchanges{with the mean shown as a black line and $1\sigma$ values shown as blue lines.}}
    \label{fig:mes_err_plot}
\end{figure}

These ten observations were processed with the same method, which produces a set of output maps each for the different parameters. By comparing the output maps to the ideal input maps, we can evaluate the measurement error for each parameter. This is shown in Figure~\ref{fig:mes_err_plot}, in which the difference between the input and output map is plotted as a function of the distance to the center for all ten observations. This plot allows us to outline the radial dependence of the measurement error for both the centroid shift and the line broadening. In our model, we then incorporate the measurement error on the centroid shift, $\Delta C_{\mathrm{mes}}$, and line broadening as such:
\begin{equation}
    \Delta C_{\mathrm{mes}} = \mathcal{N}(\mu(r), \sigma(r))
,\end{equation}
with $\mu(r)$ and $\sigma(r)$ taking the nearest value of the mean and standard deviation in each radial bin shown in Fig.~\ref{fig:mes_err_plot}. We proceeded in the same fashion for the line broadening, except that it is drawn from a truncated normal distribution, with bounds $\{0; +\infty\}$. \\

\indent One might be compelled to compare this measurement error with the errors presented on the broadening measurement with the Resolve instrument on board XRISM. For example, in \citetalias{audard_bulk_2025}, errors on the centroid shift and broadening were reported to be on the order of 1-10~km/s. The error that is reported from the XRISM measurements represents the estimated error made in fitting the centroid shift and broadening from a model to the observed spectrum. It also accounts for the calibration of the instrument.\\
The error that we report is not estimated from the spectral fitting procedure. Our measurement error is estimated from the spread of the difference between 1) the emission weighted projection of the spectral properties of each point in the cluster, and 2) the observed output maps. Because the output maps are obtained with spectral fitting, the measurement error naturally includes the uncertainties in the fitting of the spectra. The measurement error thus represents the error that is made in representing the output maps by the input model. The observed measurement error obtained in Figure \ref{fig:mes_err_plot} is then naturally explained. In larger spatial bins, far from the cluster center, the mixing of different spatial regions leads to more variation in the observed centroid shift and line broadening from the spectral fitting when compared to an emission weighted value. In the central bins, the observed centroid shift, which is the emission weighted average of the velocity cube, is driven by the peaked emission profile, i.e., a limited spatial region in the center. Because the centroid shift is sensitive to the representativeness of the projection with respect to the value obtained from spectral fitting, a bias can \boldchanges{be} observed in the \boldchanges{innermost annulus. This bias is observed as an underestimation of the measured value compared to the input expectation. The broadening is not as sensitive to the projection effects and shows no similar bias.} 

\indent \boldchanges{The measurement error that we report here is derived from the comparison between an idealized, known input and ten simulated observations. If this estimation is fit for our purpose, as it stands it does not allow us to draw any conclusions for actual measurements such as those provided by XRISM. A systematic study would be needed in order to assess any projection effects and biases that could arise from the complex spectral fitting process. Such an investigation is outside the scope of the present paper.} 
\subsection{Validation of the trained density estimators}
\label{sec:validation}

\begin{figure*}[ht]
   \includegraphics[width = 18cm]{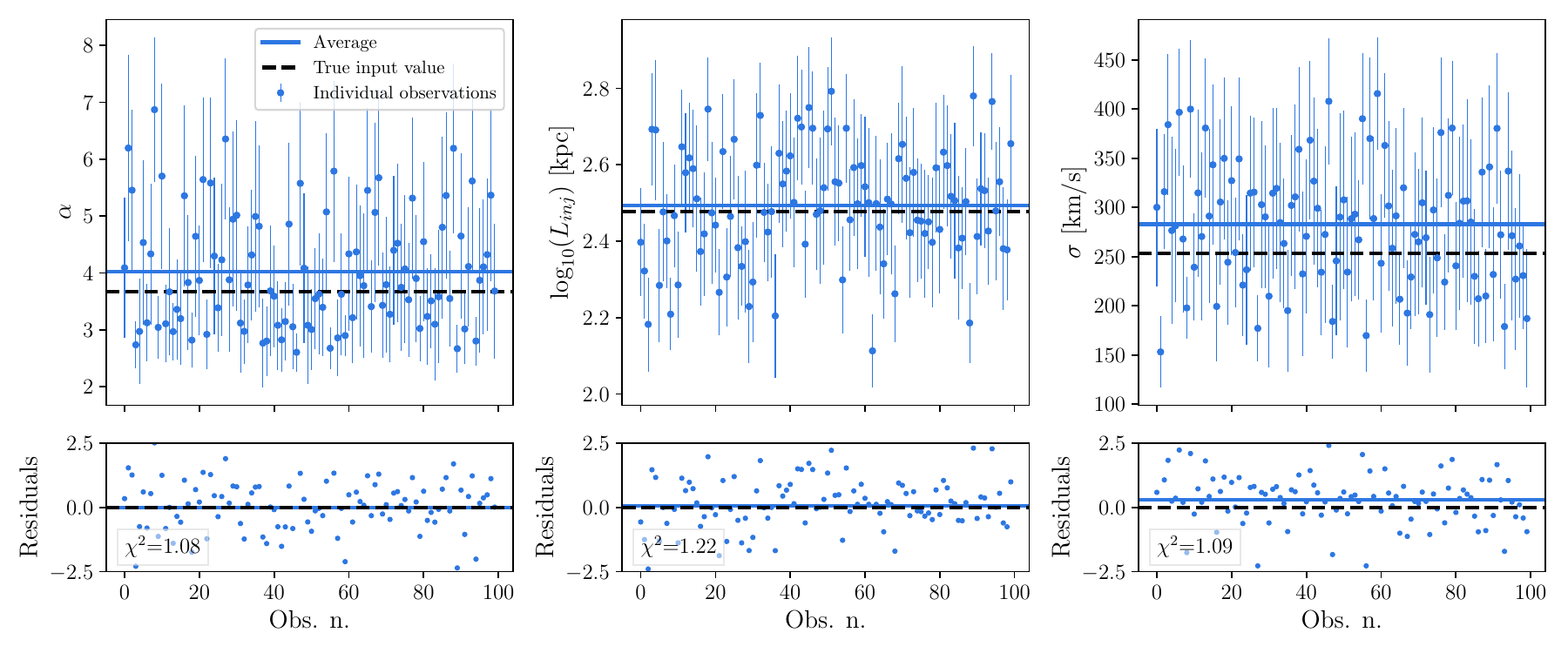}
   \caption{Retrieved average value and standard deviation of the marginalized posterior distribution of each parameter, obtained for 100 observations of 100 velocity field realizations generated with the same input parameters. The density estimator used SNPE with default parameters and was trained using the structure function of the centroid shift and the broadening.}
              \label{fig:validation_plot_100_obs}%
\end{figure*}

\indent In order to validate the trained neural density estimators, we used mock observations, each produced with the model used for the generation of training data, where each observation represents a single realization of the velocity field. The input parameters were set to the values mentioned in Sec.~\ref{sec:2}, used for the \SIXTE~ simulation. For each mock observation, we retrieved the posterior distribution, and computed its mean and standard deviation. Figure \ref{fig:validation_plot_100_obs} shows the distributions of the retrieved parameters, summarized as a single mean and standard deviation for each mock observation. This is shown here for the density estimator that was trained on the structure functions of the centroid shift and line broadening. On average, the parameters are well retrieved, despite some offset on the power spectrum normalization parameter, $\sigma$, where the averages of the samples retrieved from SBI are biased on average toward larger values. Nevertheless, when we account for the spread of the distributions by computing the residuals, they are centered on zero. In addition to that, the reduced $\chi^2$ is close to 1, with a value of 1.22 in the worst case. 
These elements show the good reconstruction of the input parameters and validate the ability of our methodology to retrieve the input parameters with the structure function. This analysis was carried out for each trained model, showing similar, or better, results. In particular, the addition of other summary statistics shows even better reconstruction.

\begin{figure}
    \centering
    \includegraphics[width=1.0\linewidth]{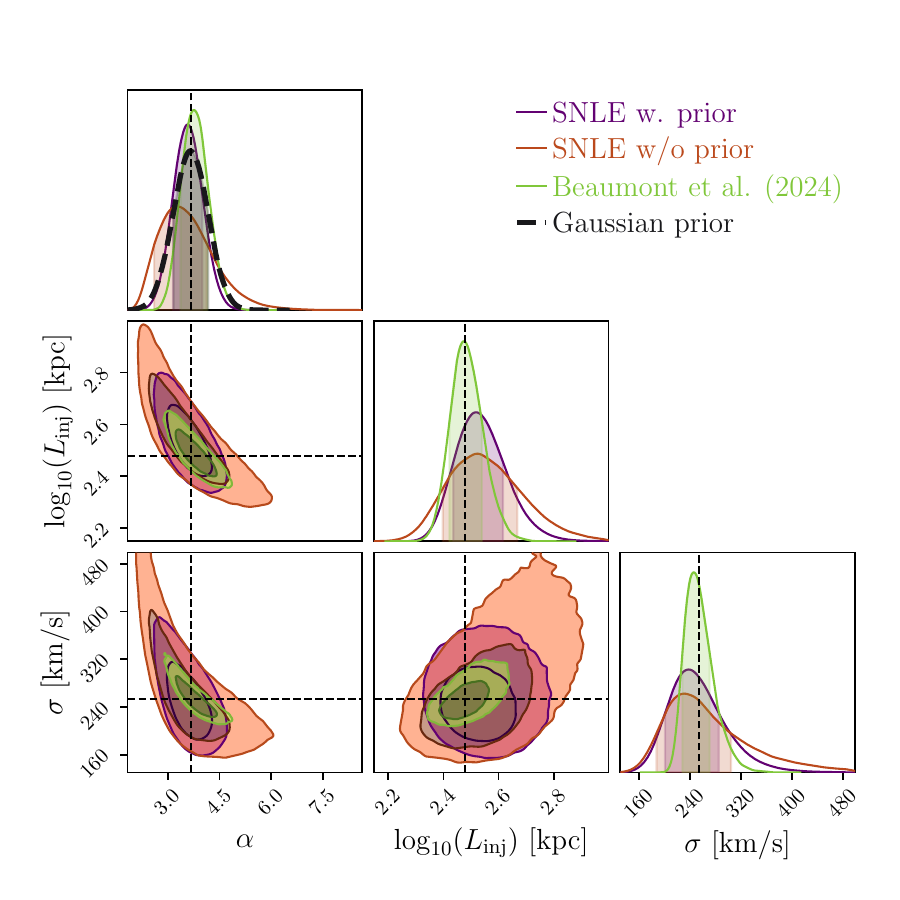}
    \caption{Posterior distributions of the parameters of the turbulent power spectrum estimated with SBI on the observation used in \cite{2024A&A...686A..41B}. A Gaussian prior has been used for the slope, $\alpha$, and is plotted with the black curve. The input values of the observation are plotted with dotted black lines.}
    \label{fig:compare_prev_paper}
\end{figure}

\indent In addition to validating the reconstruction of the parameters, we used this new framework to compare posterior distributions obtained from the observations presented in the previous paper \cite{2024A&A...686A..41B}. The result is shown in Figure \ref{fig:compare_prev_paper}, where we compare the results obtained on the 19 pointings 125~ks observation mosaic using the X-IFU specification before reformulation. In concordance with our previous paper, we used the same prior for the slope of the power spectrum, $\alpha$, and deliberately fixed the dissipation scale to its input value. We used the SNLE model for convenience, as it gives access to the likelihood and the addition of a prior on the parameters is thus straightforward. We also show the posterior distribution of the SNLE density estimator used without any prior other than what was used for the training set (as is described above). The posterior distributions obtained from SBI are larger than those obtained from the analytical method. This underlines that the approximations made in B24, in particular, neglecting the covariance of the structure function bins, are likely to have produced an underestimation of the errors, resulting in narrower distributions for the inferred parameters. 
However, the same degeneracy between the parameters can still be observed, giving us confidence that the general statistics of the problem captured by the trained neural networks follows the same underlying physics as the analytical description.

\section{Results}
\label{sec:results}
\subsection{Using the structure functions}

\begin{figure}
    \centering
    \includegraphics[width=1.0\linewidth]{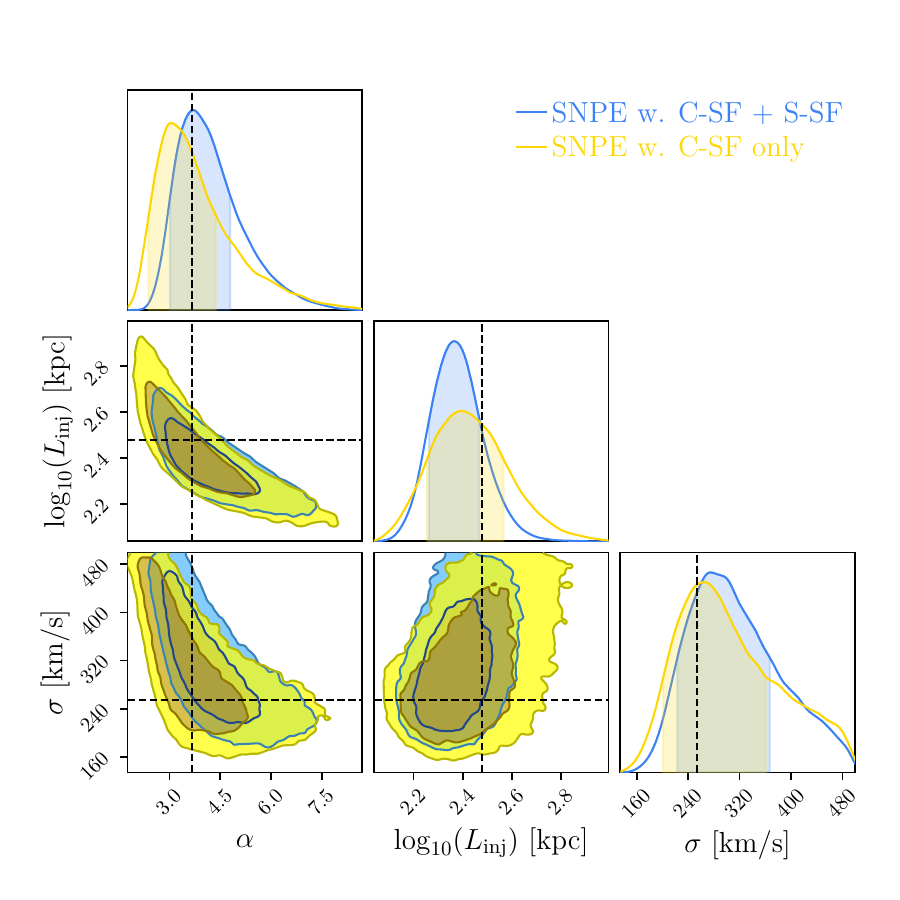}
    \caption{Posterior distributions of the parameters of the turbulent power spectrum estimated with SBI on the reference realization. The blue posterior shows the result for the density estimator trained with both the centroid shift and broadening structure functions (C-SF and S-SF). The yellow posterior shows the result of the density estimator trained only with the centroid shift structure function. The input values of the simulation are plotted with dotted black lines.}
    \label{fig:compare_broad_no_broad}
\end{figure}

\indent In Fig.~\ref{fig:compare_broad_no_broad} we show the posterior distribution obtained with training our inference network with the structure functions. We compare the result of using uniquely the centroid shift structure function, on one side, with that of using the centroid shift together with the broadening structure functions, on the other side. The two posterior distributions obtained are comparable in shape and extent. This is a result of structure function of the broadening being dominated by the measurement error. This can be seen in the difference between the input and output broadening maps in Figures~\ref{fig:input_maps} and \ref{fig:output_maps}, where the structure seen in the input map cannot be distinguished in the output map because of the fluctuation in each spatial region.  

\subsection{With additional summary statistics}
\label{sec:add_stats}
\begin{figure}
    \centering
    \includegraphics[width=1.0\linewidth]{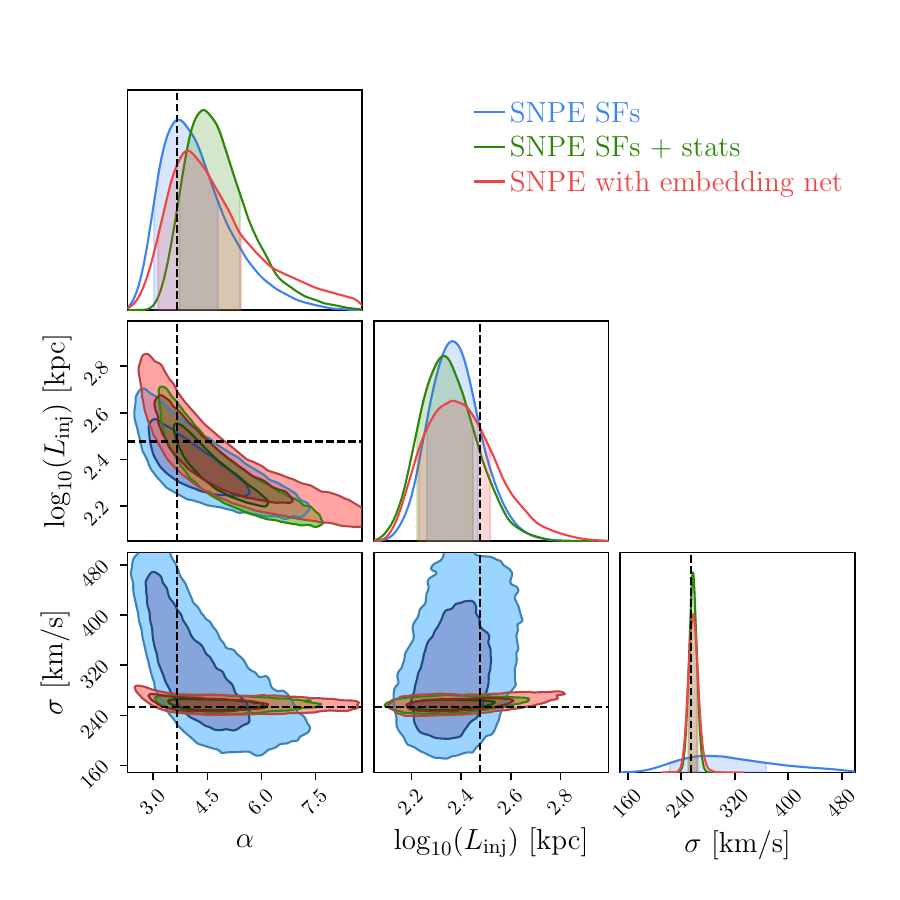}
    \caption{Posterior distributions of the parameters of the turbulent power spectrum estimated with SBI on the reference realization with different summary statistics. The input values are plotted with dotted black lines. The blue posterior was obtained with SNPE trained uniquely on the structure functions. The green posterior was obtained with SNPE trained on the structure functions together with additional statistics of the output maps. The red posterior was obtained with SNPE trained on the raw output maps with an embedding network.}
    \label{fig:sf_stats_embed}
\end{figure}

\indent Simulation-based inference allows for the addition of summary statistics in a straightforward manner. It allows us to use basic summary statistics in addition to the structure function, which is already a good description of the spatial structure of the observed maps, and hence of the structure of the ICM velocity field. If we write $C_i$, the value of the centroid shift in the bin, $i$, of the map, and $B_i$, the value of the broadening, then the statistics we used can be written as $\{ \overline{C_i}, \sigma(C_i), \overline{B_i}, \sigma(B_i) \} $, corresponding to the mean and standard deviation of the binned values obtained in each map. The resulting posterior distribution is shown in Fig.~\ref{fig:sf_stats_embed}. The high correlation between the average value of the broadening vector and the normalization of the power spectrum, $\sigma$, is captured well in the tight constraint obtained on the latter.\\
\indent In addition, we tested the use of an embedding neural network on the “raw” data, i.e., on the binned values of the output maps, represented as vectors. In other words, instead of computing the structure function from the output maps, we trained a neural network to extract significant features directly from the binned maps. This neural network was made available by the \texttt{sbi} package and was trained together with the neural density estimator. For its architecture, we chose a fully connected network\footnote{one of the architectures available in the \texttt{sbi} package, the other architectures being best adapted for extracting features out of images or out of sequential data such as time series} as it is the most relevant one to capture the covariances between the different bins of the data. The architecture was chosen as follows: the input layer has dimension $N$, with $N$ the size of the input vectors, while the output layer has dimension 16 as it is the power of two closest to the size of the structure function (in our case 20 points), allowing for a summary of the vectors in a few prominent features. We chose to have four layers, as a compromise between exceeding the minimum layers (input + output) and adding more layers, which increased the computation time during training. Changing these parameters did not result in any particular improvement in the performance of the network. The training of the embedding network was done together with the training of the neural density estimator. The resulting posterior distribution after sampling from the density estimator on the reference realization is given in Fig.~\ref{fig:sf_stats_embed}. It is comparable to the posterior obtained with SNLE trained on the structure functions and additional statistics. In particular, the embedding network also allowed the extraction of a better constraint on $\sigma$ compared to using only the structure functions. The embedding network is therefore unable to guide our intuition toward additional summary statistics that could improve the inference, and only achieves at best similar constraints to those obtained from the density estimator trained with the structure function and statistics.\\

\begin{figure}
    \centering
    \includegraphics[width=1.0\linewidth]{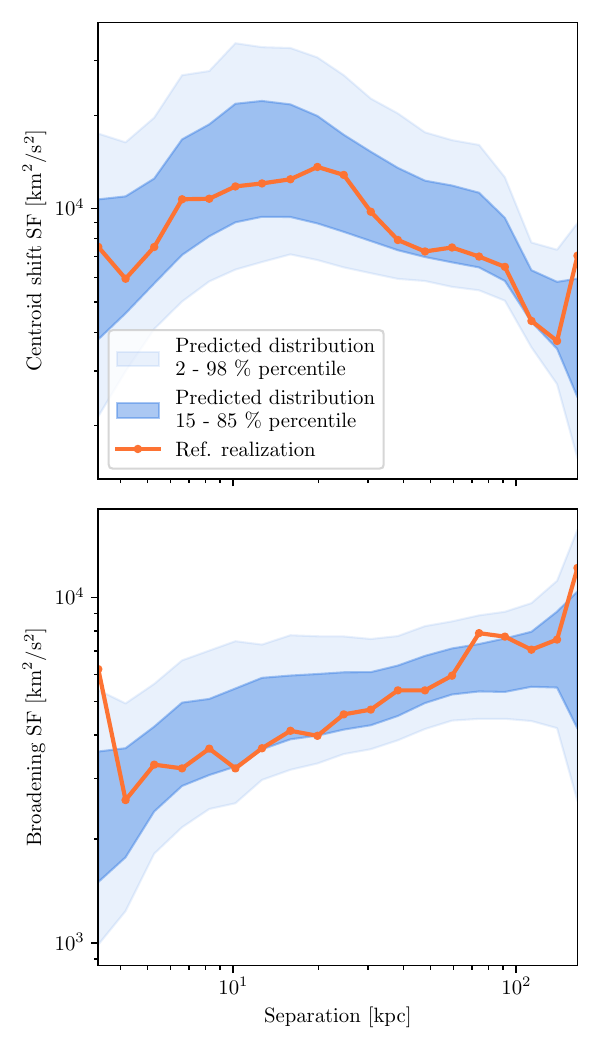}
    \caption{Posterior predictive check for the model SNPE trained with the structure function and additional statistics, evaluated on the reference realization. It shows the predicted distribution of the structure functions of the centroid shift and line broadening distributions as envelopes, with parameters sampled from the posterior distribution. The reference realization is plotted in orange and is the same as Fig.~\ref{fig:pred_sf_dist}.}
    \label{fig:ppc_sf_plus_stats}
\end{figure}

\begin{figure}
    \centering
    \includegraphics[width=1.0\linewidth]{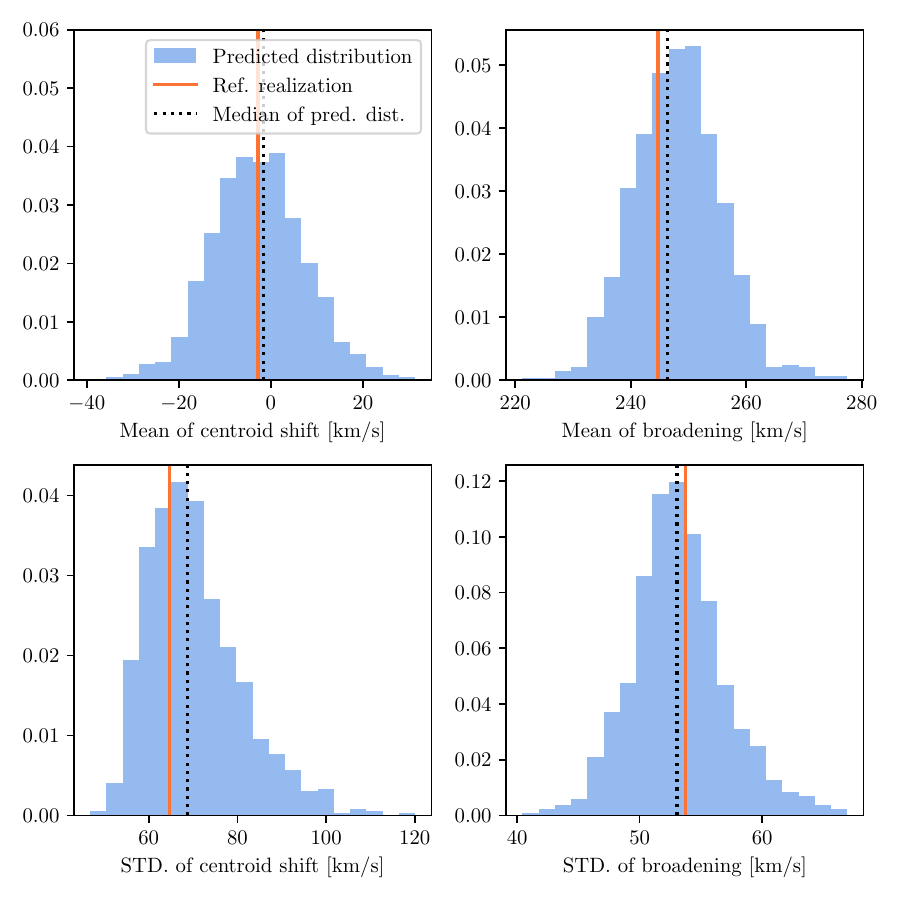}
    \caption{Posterior predictive check for the model SNPE trained with the structure function and additional statistics, evaluated on the reference realization. It shows the predicted distribution of averages and standard deviations of the binned values of the centroid shift and broadening. \boldchanges{The median value of each distribution is plotted as a dotted black line.} The reference realization is plotted in orange.}
    \label{fig:ppc_sf_plus_stats_2}
\end{figure}

\indent To complement the validation presented in Sec.~\ref{sec:validation}, we used the posterior predictive check. It consists of running the model on parameter samples drawn from the posterior distribution, and comparing the output distribution with the observable used for the inference. \boldchanges{For this check, we used 1000 samples drawn from the posterior distribution.} In Fig.~\ref{fig:ppc_sf_plus_stats}, we show the result of the posterior predictive check for the model “SNPE SF + stats” with the structure function. In Fig.~\ref{fig:ppc_sf_plus_stats_2} we show the result with the additional statistics used. The distributions obtained from the posterior predictive check show a good recovery of the properties of the reference realization. \boldchanges{The medians of the predicted distributions are close to the reference observation properties. The non-Gaussian shape of the distributions is explained by two facts. Firstly the posterior distribution of the parameters is not a multivariate Gaussian, and secondly the mapping between the space of posterior parameters and observed properties is not straightforward, so there is no reason for the distribution of the predictive posterior check to be perfectly Gaussian.} 

\subsection{Summary}

\begin{figure}
    \centering
    \includegraphics[width=0.95\linewidth]{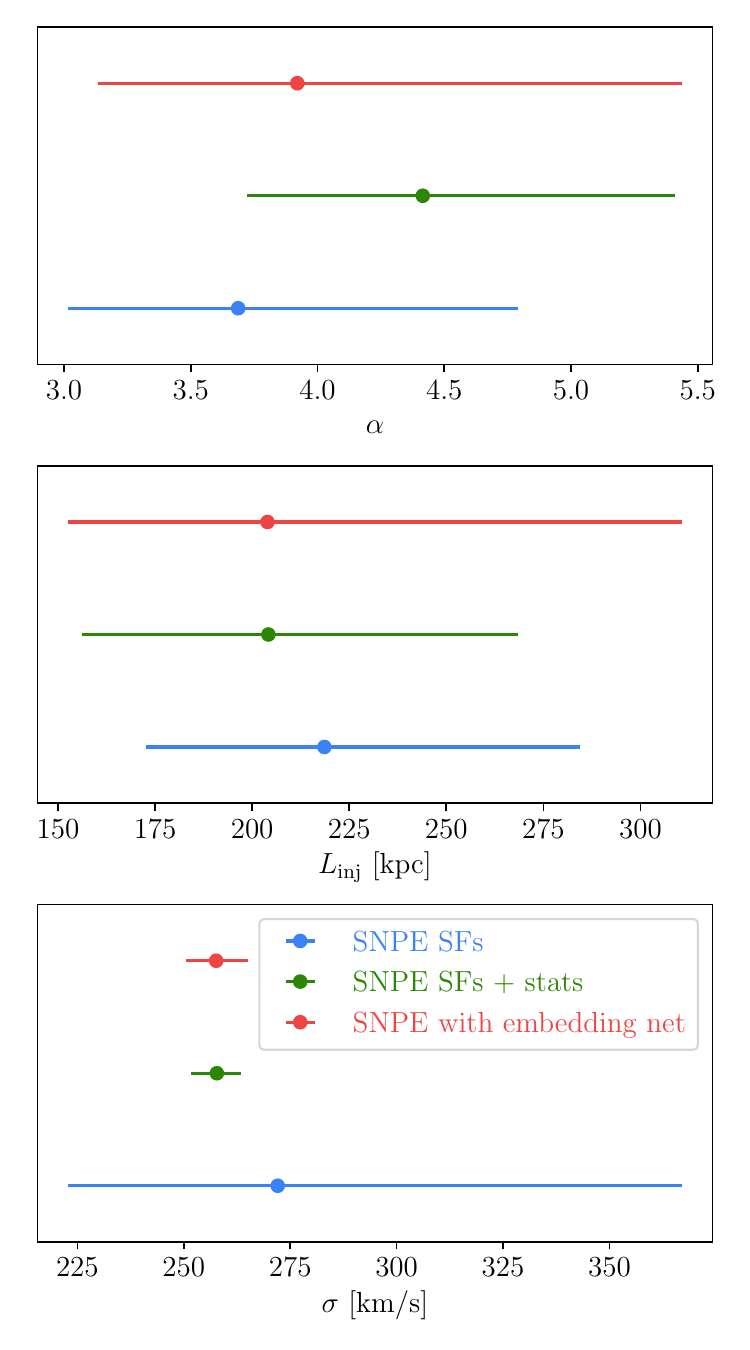}
    \caption{Summary plot of the average values and $\pm1\sigma$ errors of marginalized distributions of each parameter for different models, obtained on the reference realization. The true input parameters are plotted as a vertical dotted line. The depicted errors include the full budget of the inference uncertainty including sample variance and measurement errors.}
    \label{fig:model_summary}
\end{figure}

\indent In Fig. \ref{fig:model_summary}, we summarized the posterior distributions obtained for each model with its average value and standard deviation. None of the models are able to put a constraint on $\alpha$ better than approximately $\pm1$, showing the difficulty in retrieving this parameter from such observations. The precision on the injection scale, $L_\mathrm{inj}$, is approximately $-50,+100$kpc; however, for this particular realization, the retrieved injection scale is biased toward values lower than the true input. This order of magnitude for the precision of the injection scale would allow different turbulence driving mechanisms (AGN feedback, sloshing, structure assembly) to be distinguished between. The norm of the power spectrum, $\sigma$, is the parameter that is best retrieved, with a constraint of $\pm5$~km/s with the SNPE trained on the structure functions with additional statistics. This error connects directly to the precision of the estimated nonthermal pressure support. The offsets between these average values and the true input values is explained by the cosmic variance, as is shown in Fig.~\ref{fig:validation_plot_100_obs}, in which any particular realization can exhibit properties that lead the inference to obtain parameters that are offset from the true values. \\

\section{Discussion}
\label{sec:discusssion}

\subsection{Limitations of the model hypotheses}

\indent The toy model used for simulating  the galaxy cluster considered in this study is an approximation, which assumes spherical symmetry with no departure from the ideal thermodynamical profiles. A realistic simulation would need to account for such effects and their statistics across a galaxy cluster population. Moreover, the parameters of the turbulent field are expected to vary within individual galaxy clusters \citep{dupourque_chex-mate_2024, eckert_non-thermal_2019}. Then, modeling the turbulent field as a single homogeneous and isotropic field with a single set of turbulent parameters is an approximation that would need to be refined to capture the complex dynamics and substructures present in galaxy clusters. This is likely to considerably increase the complexity of the reconstruction of turbulence power spectrum parameters  in a single cluster.

\indent In the present paper, we used \texttt{xspec} for fitting the spectra in the different spatial regions. In order to facilitate the fitting procedure, we binned the data in spatial regions of an equal signal-to-noise ratio of 200, using Voronoi binning. Using Bayesian methods for fitting \citep[such as][]{buchner_x-ray_2014, dupourque_jaxspec_2024, 2024A&A...686A.133B} would allow us to greatly reduce the size of the spatial regions, with no need to rely on such a high signal-to-noise ratio, and would allow us to make use of the uncertainty in the inferred spectral parameters. Leveraging surrogate models such as \texttt{SUSHI} \citep{lascar_sushi_2024} could also allow us to retrieve the spatial spectral properties of the emission to a higher degree of spatial precision, without sacrificing the resolution. These methods may be investigated in a future paper, and should be noted as existing alternatives to the current method.

\indent In the present study, extended work was required to ensure that the model used for producing the training data was comparable with the outputs of the \SIXTE~ simulations. In particular, the measurement error is fine-tuned to correspond to a particular realization, which has two shortcomings. Firstly, this measurement error represents the error with respect to a “true” known input, which is only possible in simulations such as ours. Secondly, it requires one to generate a training set for SBI for each realization that serves as a reference. Ideally, one would have a training set that is valid for any realization of any velocity cube. In order to achieve that, one would reproduce the result shown in Fig.~\ref{fig:mes_err_plot} for different velocity cubes simulated within a cluster observed with X-IFU through \SIXTE. Then, the statistics of the measurement error could be generalized for all velocity cubes.

\subsection{Impact of background}

\begin{figure}
    \centering
    \includegraphics[width=1.0\linewidth]{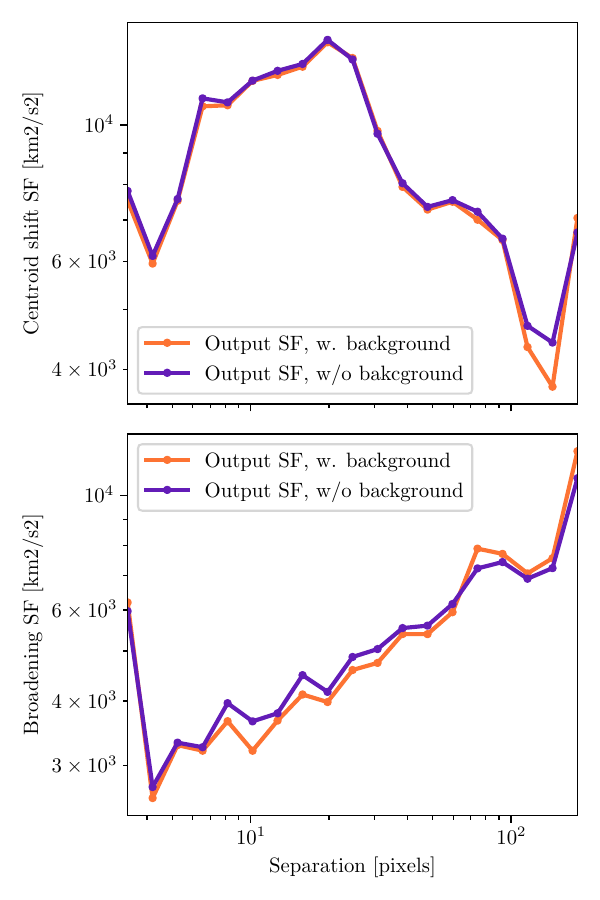}
    \caption{Structure functions of the centroid shift (top) and broadening (bottom)  maps, produced for a reference observation. The orange curve shows the observation including background, the purple curve shows the observation with the background excluded from the modeling.}
    \label{fig:compare_sfs_bkg}
\end{figure}

\begin{figure}
    \centering
    \includegraphics[width=1.0\linewidth]{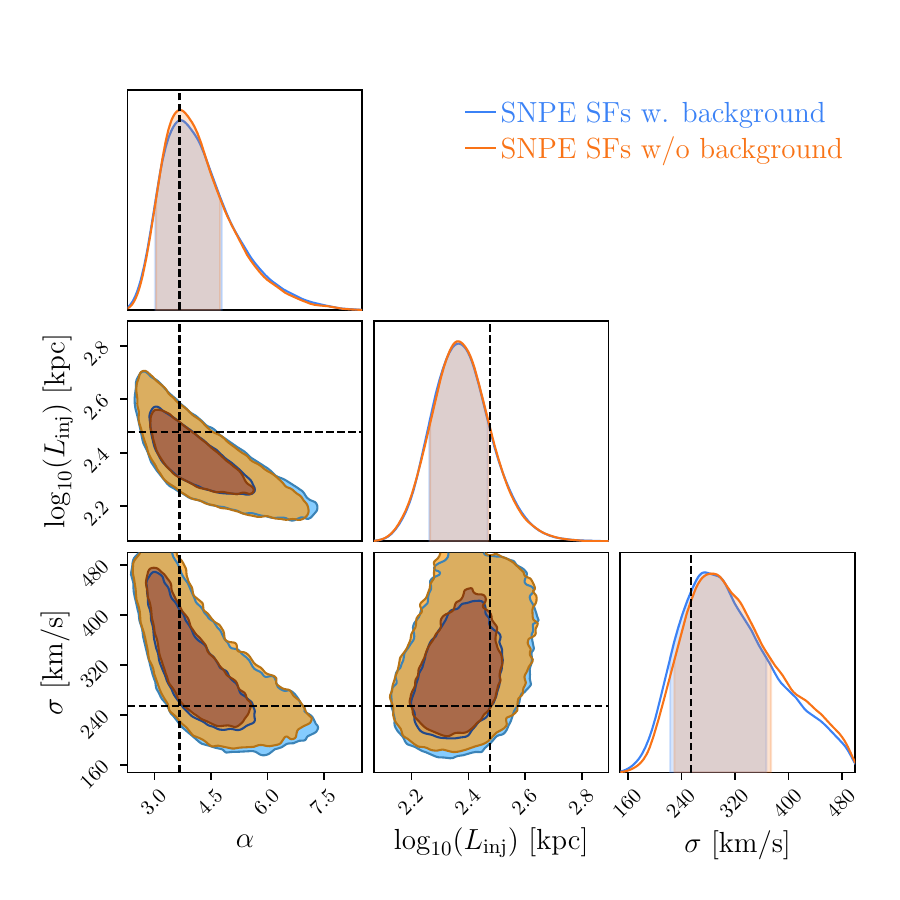}
    \caption{Posterior distributions of the parameters of the turbulent power spectrum estimated with SBI on the reference realization. The blue posterior shows the result of the density estimator evaluated on the realization that included background. The orange posterior shows the result of the density estimator evaluated on the realization which included no background. The input values of the realization are plotted with dotted black lines.}
    \label{fig:compare_bkg_no_bkg}
\end{figure}

\indent The impact of the astrophysical background has been assessed to be negligible for the measurement of turbulence for the toy model cluster simulated in the present work. It should be noted that the level of the astrophysical background is low compared to the instrumental background. The \SIXTE~ simulator allows a flagging of the events that are sampled from the background. By excluding these events from the generation of spectra and producing observed centroid shift and broadening maps for the reference realization, we are able to reproduce the observation of the reference realization in the absence of background. The resulting structure functions are shown in Fig.~\ref{fig:compare_sfs_bkg}. This shows that the background has a negligible effect on the retrieved structure functions of the centroid shift and line broadening. The obtained posterior distribution with and without background are shown in Fig.~\ref{fig:compare_bkg_no_bkg}, and are identical. However, we only simulated a uniform diffuse background, neglecting the presence of any point-like source in the FoV. Indeed, a more complete approach would be to sample AGNs from a population, include them in the observation, mask all those that are resolved, and treat the unresolved component as a diffuse background \citep[see e.g.,][]{cucchetti_towards_2019, castellani_case_2024}. Nevertheless, the main differences to our approach would be twofold. Firstly, masked sources would appear in the mosaic of pointings, which would have only a little impact on the observable properties of turbulence. Masking point sources only reduces the signal-to-noise by reducing the total area from which the structure function is computed. Secondly, the sources that might have not been properly masked would show as additional residual fluctuations in the CXB.


\subsection{Sample variance}
\label{sec:sample_variance}

\indent Our analysis demonstrates the full potential of SBI-based methods for doing inference on problems where the likelihood is either ill defined, difficult to express analytically, or costly to evaluate numerically. In the context of evaluating turbulence in galaxy clusters, the structure function of the maps of the centroid shift and broadening, as computed in our analysis, is subject to a high level of sample variance. That is, a single set of turbulent parameters can give rise to a wide range of observed velocity fields and associated structure functions, due to the fact that the underlying turbulent field is stochastic and acting at scales comparable to that of the cluster itself. This was already outlined in \citet{2024A&A...686A..41B}, and is emphasized again in our present paper. The analytical description of the contribution of cosmic variance is a complex task, which can lead to an underestimation of the full error budget despite careful accounting due to necessary  approximations, in particular, neglecting the covariance of the structure function bins to make the computation tractable. The posterior retrieved by sampling the analytical likelihood is much narrower than the posterior obtained from using SBI, showing how SBI fully encapsulates the total variance of the structure function, resulting in wider inferred distributions. \\

\indent In this paper, we only presented results for the “reference realization,” which was picked as its structure function lies close to the average predicted structure functions for the set of input parameters, in order to facilitate the interpretation of the results. To provide a more complete view of the results of inference on different realizations, in Appendix \ref{AppendixB} we show the posterior distributions obtained with the same set of models on two different realizations for which the structure function lies, respectively, higher (on average, the centroid shift SF is 1.9$\sigma$ above the mean predicted  centroid shift SF, see Fig.~\ref{fig:posteriors_obs3_obs4}, left panel) and lower (on average, the centroid shift SF is 0.7$\sigma$ below the mean predicted centroid shift SF, see Fig.~\ref{fig:posteriors_obs3_obs4}, right panel). This is in line with the results shown in Fig.~\ref{fig:validation_plot_100_obs}, which shows that the high level of sample variance causes a large spread in the retrieved distributions with respect to the true input values. This further emphasizes the importance of accounting for the impact of the sample variance when retrieving turbulence parameters. This effect could be mitigated by averaging observations of several targets, considering each of them as  independent observations, i.e., realizations, of the same stochastic physical process.

\subsection{Comparison to existing work}

\indent \boldchanges{Besides our previous work, another feasibility study of X-IFU abilities to measure turbulence in galaxy clusters has been conducted by \citet[][R18 hereafter]{roncarelli_measuring_2018}. The authors used a Coma-like hydrodynamical simulation of a galaxy cluster at $z=0.1$ including physically implemented turbulence. They simulated a single X-IFU pointing at the cluster center for an exposure of $\sim2$~Ms, allowing them to measure the structure function for different Mach numbers of the parent hydro-simulation, at low separations and with low statistical errors. The higher statistics due to their exposure time over a single pointing demonstrated the ultimate capability of X-IFU, such as mapping physical properties at the instrument pixel level. In contrast, our 125~ks exposure time per pointing provides a much more realistic view of expected pointings with X-IFU, but at the expense of the statistical precision. R18's preliminary work, however, was performed with an early instrumental configuration of NewAthena/X-IFU, whereas our current study used the latest performance parameters for NewAthena/X-IFU. As was acknowledged in R18, a single pointing limits the range of accessible scales needed to fully characterize the turbulent cascade. Moreover they also did not account for the stochastic nature of the turbulence and the induced sample variance, whereas we demonstrate in this study (Sec.~\ref{sec:sample_variance}) and throughout our series of papers the key impact on the total error budget when measuring turbulent velocities. Properly accounting for this error component will be instrumental for our actual ability to extract and interpret the role of ICM turbulence in the formation and the evolution of the galaxy cluster population. The work of \citet{Zhang_LEM_2024} must also be mentioned in investigating the capabilities of the Line Emission Mapper (LEM) proposed to NASA as a high-resolution integral field unit focused on low-energy observations. In particular, they are able to show its ability to measure the injection scale from a 1~Ms simulated observation of a hydrodynamical simulation, by using the second-order structure function of the velocity field. A key development in this study is a new spectral model enabling Gaussian-distributed emission measures while generating velocity-broadened lines. It thus partially addresses those problematic projection effects along the line of sight.}

\indent \boldchanges{On the side of observations, the recent measurements from the Resolve instrument on board XRISM are paving the way for direct measurements of gas motions in galaxy clusters. The reported nonthermal pressure supports derived from the observed spectral lines shifts and broadening are on the order of $\sim1-4$~\%, suggesting lower levels of turbulence than are expected from numerical simulations \citepalias{audard_bulk_2025, xrism_collaboration_xrism_2025, xrism_collaboration_2025_abell_2029_a}. Although Resolve has an exquisite spectral resolution (better than 5~eV) and is obtaining transformational measurements, the spatial mixing induced by the PSF ($\sim 1$~arcmin full width at half maximum), the limited number of pixels paving the FoV (35 active pixels over $3\times 3$~arcmin$^2$) and the relatively low effective area of its telescope \citep{10.1117/12.3018882} limit XRISM's ability to map spectral features or cover larger scales, and allow only access to nearby clusters. As a consequence, this also restricts the current constraints produced on structure functions to a few separations, leading to an inherent very high sample variance  \citepalias{xrism_collaboration_xrism_2025}. However, these measurements pave the way for future observation with Athena/X-IFU as illustrated in the present work. Results from XRISM will drive future approaches. For instance, if low levels of turbulence were to be confirmed over a larger sample of galaxy clusters, deeper feasibility studies with X-IFU would need to account for it. To that end, we would need to devise the best observing and analysis strategies to extract line centroid shift and broadening to further fathom the role of turbulence in galaxy clusters.}

\section{Conclusions}

\indent In this paper, we have investigated the use of SBI for the retrieval of turbulent power spectrum parameters from X-ray observations of galaxy clusters with the X-IFU instrument on board the NewAthena space observatory. We followed the framework introduced in the previous installment of this series of papers, with the simulation of a mosaic of 19 X-IFU pointings, updated to the current X-IFU instrument configuration. We can summarize our work with the following points:
\begin{itemize}
    \item We performed end-to-end simulations of an observation of a $z=0.1$ galaxy cluster with a mosaic of $19\times125$ks pointings of the X-IFU instrument, using a physically motivated thermodynamical toy model and underlying turbulent velocity field. 
    \item We developed a model for generating observed centroid shift and line broadening maps rapidly, and used this model for generating a large amount of training samples. We used these samples to train the SNPE model as implemented in the \texttt{sbi} package, which is a neural network trained to map the correspondence between distributions of observable properties and the parameters used for the generation of these observables. This method naturally encompasses all sources of variance, and in particular the sample variance arising from the stochastic nature of the turbulent velocity field. Because the injection scale of turbulence we modeled is on the order of the typical radius of the emissivity field, the sample variance has a notable impact on the inference of the turbulent of parameters, which needs to be accounted for. 
    \item We showed that the trained networks were able to retrieve the turbulent parameters of the underlying field with reasonable accuracy (see Fig.~\ref{fig:validation_plot_100_obs}). In particular, we were able to compare the posterior distribution issued from this method to the one issued from sampling the analytical likelihood, as was done in \cite{2024A&A...686A..41B}. The posterior distribution obtained with SBI is much larger (see Fig.~\ref{fig:compare_prev_paper}), as it accounts for the full error budget including the sample variance inherent to the physical nature of turbulence, whereas, in the analytical model, it is only partially captured, leading to underestimated errors.
    \item We conducted inference by using the structure function as a summary statistic of the centroid shift and broadening maps, and we also explored the use of additional summary statistics. Notably, the use of the structure functions jointly with the average and standard deviation of the binned values within the maps has proven to be the most effective in constraining the norm of the power spectrum. We explored the use of a fully connected neural network to automatically learn relevant summary statistics from the maps. It showed a similar performance to using the structure functions with the averages and standard deviations (see Fig.~\ref{fig:sf_stats_embed}).
\end{itemize}

In a subsequent study, we shall investigate observation strategies with X-IFU, in trying to maximize the scientific output for a given total exposure. Overcoming sample variance is likely to be the most challenging aspect. This could be achieved through the observation of a sample of targets, as joining the observations of $N$ clusters should reduce the error on each parameter by a factor of $\sqrt{N}$ . Such a task will be a major endeavor for the X-IFU instrument. Finally, we might investigate the joint utilization of X-IFU and wide field imager (WFI), leveraging both direct measurements of turbulence as well as surface brightness fluctuations seen by WFI.

%
%

\begin{acknowledgements}
AM, SD, EP, FP and NC acknowledge the support of CNRS/INSU and CNES. AM thanks F. Mernier and A. Mahesh for their help and comments. This work was granted access to the HPC resources of CALMIP supercomputing center under the allocation 2016-P22052. The following python packages have been used throughout this work : \texttt{astropy} \citep{astropy:2013, astropy:2018, astropy:2022}, \texttt{chainconsumer} \citep{Hinton2016}, \texttt{cmasher} \citep{2020JOSS....5.2004V}, \texttt{jax} \citep{frostig_compiling_2019}, \texttt{matplotlib} \citep{Hunter:2007} and \texttt{sbi} \citep{tejero-cantero2020sbi}. 

\end{acknowledgements}

\bibliographystyle{aa} 
\bibliography{biblio.bib}

\begin{appendix} 

\section{Appendix A} \label{AppendixA}

\indent We summarize the cluster density and temperature profiles as follows. The profiles were taken from the joint analysis of eight non-cool core clusters presented in the XMM-{\it Newton} Cluster Outskirts Project \cite[X-COP][]{Ghirardini2019}. The temperature profile is described as:
\begin{equation}
    \dfrac{T(x)}{T_{500}} = T_0 \dfrac{\frac{T_{min}}{T_0} + (\frac{x}{r_{cool}})^{a_{cool}}}{1 + (\frac{x}{r_{cool}})^{a_{cool}}} \frac{1}{\left(1 + (\frac{x}{r_t})^2\right)^{\frac{c}{2}}}
\end{equation}
where $x = r/R_{500}$, and \{$T_0, T_{min}, r_{cool} a_{cool}, r_t, c/2$\} are six temperature, radius and shape parameters, with $T_{500}$ following a parametrization on the redshift, see \cite{Ghirardini2019}. The values of these parameters are provided in Table \ref{XCOPparameters}. 

\begin{table}[h]
    \centering
    \begin{tabular}{|c|c||c|c|}
       \hline
        Parameter & Value & Parameter & Value\\
        \hline \hline
        $T_0$ & 1.09  & $\log(n_0)$ [$\text{cm}^{-3}$] & -4.9 \\
        $T_{min}/T_0$ & $0.66$& $\log(r_c)$ [kpc] & -2.7\\
        $\log(r_{cool}) $ &  -4.4 & $\log(r_s)$ [kpc] & -0.51 \\
        $a_{cool}$ & 1.33 & $\alpha$ &  0.70\\
        $r_t$ & 0.45  & $\beta$ & 0.39\\
        $c/2$ &  0.30 & $\epsilon$ & 2.60\\
        $\gamma$ & 3& &\\
        
         \hline
    \end{tabular}
    \caption{Values of the parameters used for the temperature and emission models, obtained from the best-fit of these profiles to observed clusters \citep{Ghirardini2019}.}
    \label{XCOPparameters}
\end{table}
The density profile is described as: 
\begin{equation}
    n_e^2(x)= n_0^2 \frac{(\frac{x}{r_c})^{-\alpha}}{(1 + (\frac{x}{r_c})^2)^{3\beta -\alpha /2}} \frac{1}{(1 + (\frac{x}{r_s})^{\gamma})^{\frac{\epsilon}{\gamma}}}
,\end{equation}
with x = r/$R_{500}$ and \{$\gamma,n_0, r_c, r_s,\alpha,\beta,\epsilon$\} are six shape parameters, with values provided in Table \ref{XCOPparameters}. We set an arbitrary upper limit on the electron density value at 0.05\,cm$^{-3}$, in order to avoid divergence as the center of the cluster. \\
The metallicity profile follows the shape derived by \citet{Mernier2017}, which was obtained from the study of a sample of 44 nearby cool-core clusters, groups, and ellipticals observed with XMM-{\it Newton}:
\begin{equation}
    Fe(x) = 0.21 (x + 0.021)^{-0.48} - 6.54\times \exp\left(- \frac{(x + 0.0816)^2}{0.0027}\right)
\end{equation}
with $x = r/R_{500}$. The solar abundances are fixed to those of \cite{AndersAndGrevesse89}.

\section{Appendix B} \label{AppendixB}

\begin{figure}[h!]
    \centering
    \includegraphics[width=0.7\linewidth]{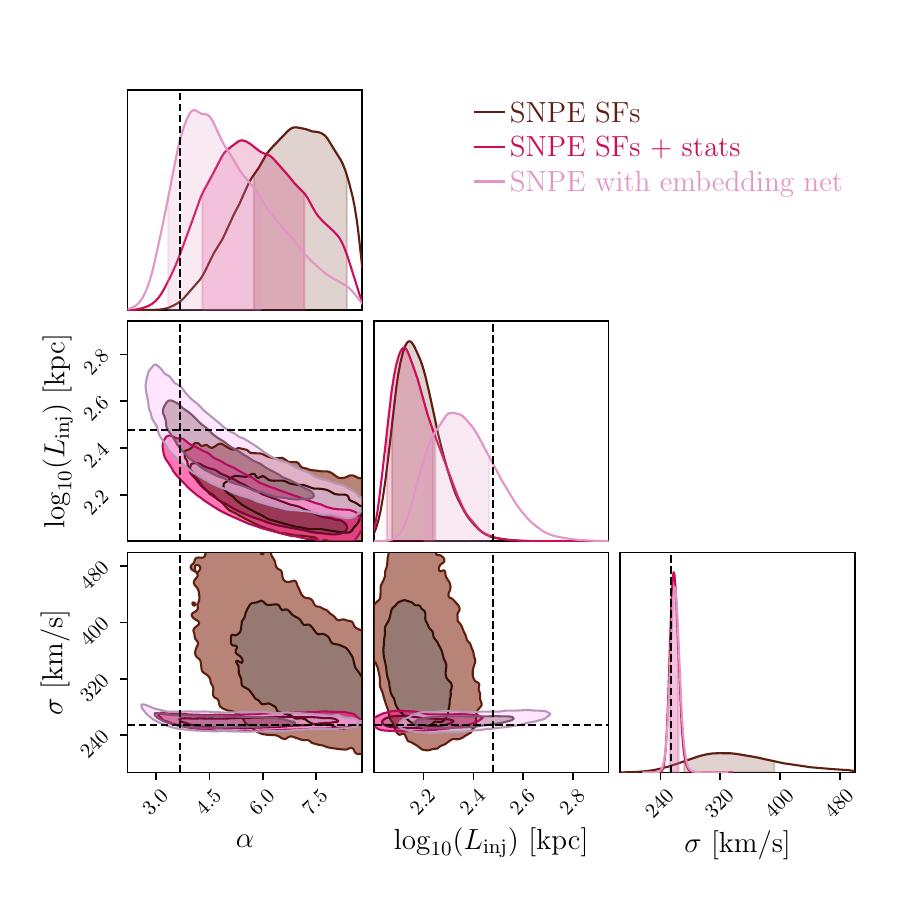} \\
    \includegraphics[width=0.7\linewidth]{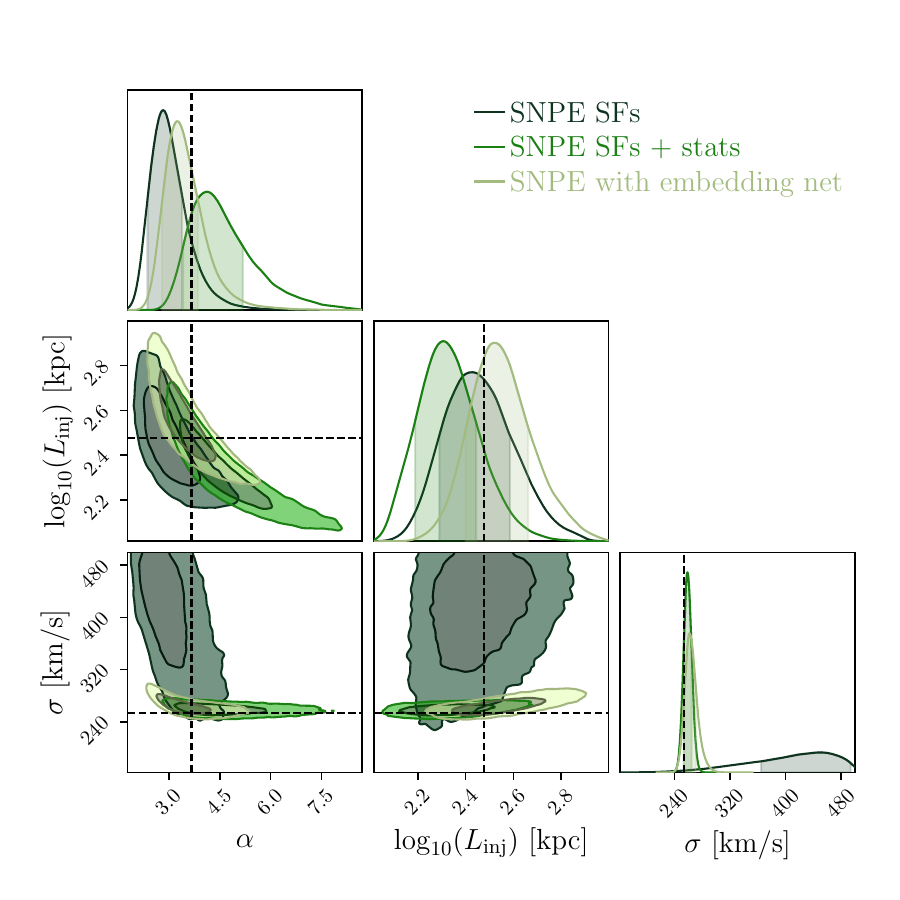}%
    \caption{Posterior distributions of the parameters of the turbulent power spectrum estimated with SBI on two realizations with different summary statistics. The input values of two realization are plotted with black dotted lines. (Top) Results for 'Obs. high' (Bottom) results for 'Obs. low'.}        
    \label{fig:posteriors_obs3_obs4}
\end{figure}

\begin{figure}
    \centering
    \includegraphics[width=1.0\linewidth]{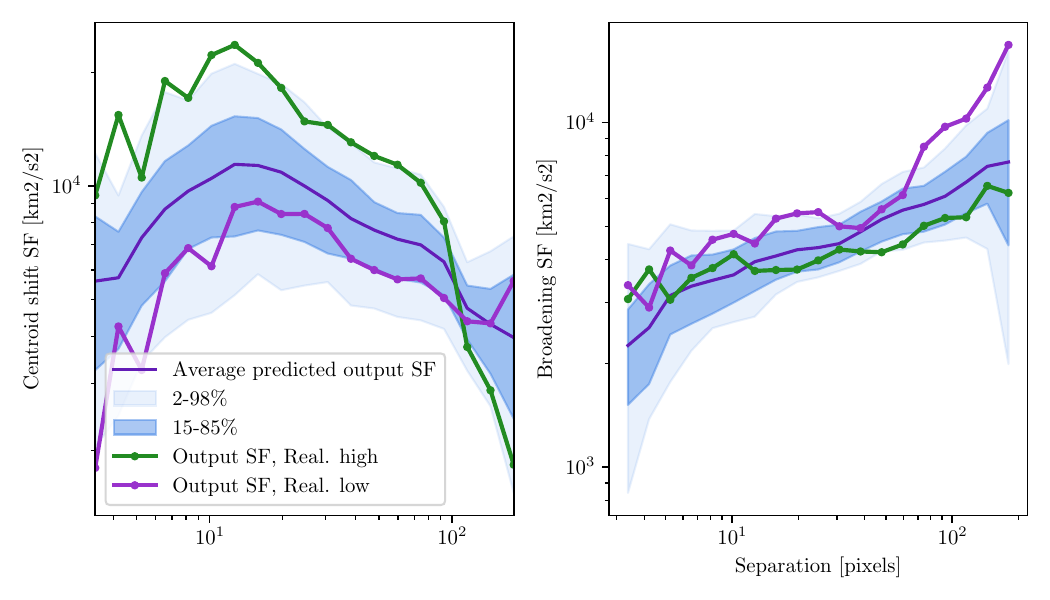}
    \caption{Posterior distributions of the parameters of the turbulent power spectrum estimated with SBI on the reference realization with different summary statistics. The input values of the realization are plotted with black dotted lines. The blue posterior was obtained with SNPE trained uniquely on the structure functions. The green posterior was obtained with SNPE trained on the structure functions together with additional statistics of the output maps. The red posterior was obtained with SNPE trained on the raw output vectors with an embedding network.}
    \label{fig:sfs_obs3_obs4}
\end{figure}

\indent In Fig.~\ref{fig:posteriors_obs3_obs4} we show the results of the models presented in the Results section of the paper on two mock realizations of a mosaic of 19 X-IFU pointings obtained by with \SIXTE, 'Real. high' and 'Real. low', the structures functions of which are shown in Fig.~\ref{fig:sfs_obs3_obs4}. 

\end{appendix}
\label{LastPage}
\end{document}